\documentclass[12pt]{iopart}

\usepackage{import} 
\usepackage{color}  
\usepackage{graphicx} 
\usepackage{braket}
\usepackage{siunitx}
\usepackage{upgreek}

\begin{document}

\title{Coulomb anti-blockade in a Rydberg dressed gas}

\author{A. D. Bounds, N. C. Jackson, R. K. Hanley, E. M. Bridge, P. Huillery, and M. P. A. Jones}

\address{Joint Quantum Centre Durham-Newcastle, Department of Physics, Durham University, Durham, DH1 3LE, UK}
\ead{a.d.bounds@durham.ac.uk}
\vspace{10pt}
\begin{indented}
\item[]December 2018
\end{indented}

\begin{abstract}
We perform a comprehensive investigation of the coupling between a Rydberg-dressed atomic gas and an ultra-cold plasma. Using simultaneous time-resolved measurements of both neutral atoms and ions, we show that plasma formation occurs via a Coulomb anti-blockade mechanism, in which background ions DC Stark shift nearby atoms into resonance at specific distances. The result is a highly correlated growth of the Rydberg population that shares some similarities with that previously observed for van der Waals interactions. We show that a rate equation model that couples the laser-driven Rydberg gas to the ultra-cold plasma via a Coulomb anti-blockade mechanism accurately reproduces both the plasma formation and its subsequent decay. Using long-lived high angular momentum states as a probe, we also find evidence of a crossover from Coulomb anti-blockade to Coulomb blockade at high density. As well as shedding light on loss mechanisms in Rydberg-dressed gases, our results open new ways to create low-entropy states in ultra-cold plasmas.
 
\end{abstract}

\section{Introduction/motivation}
Over the last 20 years, the excitation of laser-cooled atoms to high-lying Rydberg states has emerged as a powerful technique for the creation and study of strongly-interacting many-body systems. Experiments are generally carried out in the frozen gas regime \cite{Mourachko1998,Anderson1998}, where the van der Waals interaction between neighbouring Rydberg atoms dominates the kinetic energy. However, the finite lifetime of Rydberg states due to spontaneous decay and black body radiation \cite{Gallagher1995} have provided a fundamental limit to the timescale over which the system can be studied and manipulated.
	More recently, Rydberg dressing has emerged as a route to extend the interrogation time \cite{Santos2000,Bouchoule2002,Johnson2010,Henkel2010,Pupillo2010,Honer2010,Glaetzle2012}.  An off-resonant coupling laser admixes a small amount of the Rydberg state into the ground state, providing a much longer lifetime at the expense of reduced interaction strength. The many-body nature of the dressed interaction also has qualitatively new features, such as a finite or ``soft-core'' interaction potential at short range that was proposed as a route to create many-body spin squeezed states for quantum metrology \cite{Bouchoule2002,Gil2014}. The long timescales associated with Rydberg dressing also open up the possibility of coupling the strong interactions between Rydberg atoms to the motional state of the atoms, leading for example to a possible supersolid phase in Bose-Einstein condensates \cite{Henkel2010,Pupillo2010,Honer2010} . 
Experimentally, Rydberg-dressing has been demonstrated for two atoms \cite{Jau2015} and in optical lattices \cite{Zeiher2016,Zeiher2017}.	However, attempts to observe Rydberg dressing in high-density BECs and 3D lattice gases have been hampered by unexpected loss mechanisms associated with enhanced population of the Rydberg state, which have in general limited the achievable lifetime to less than 1 ms \cite{Aman2016,Goldschmidt2016,Helmrich2016,Boulier2017}.

\begin{figure}[h]
\centering
\includegraphics[width=14.4cm]{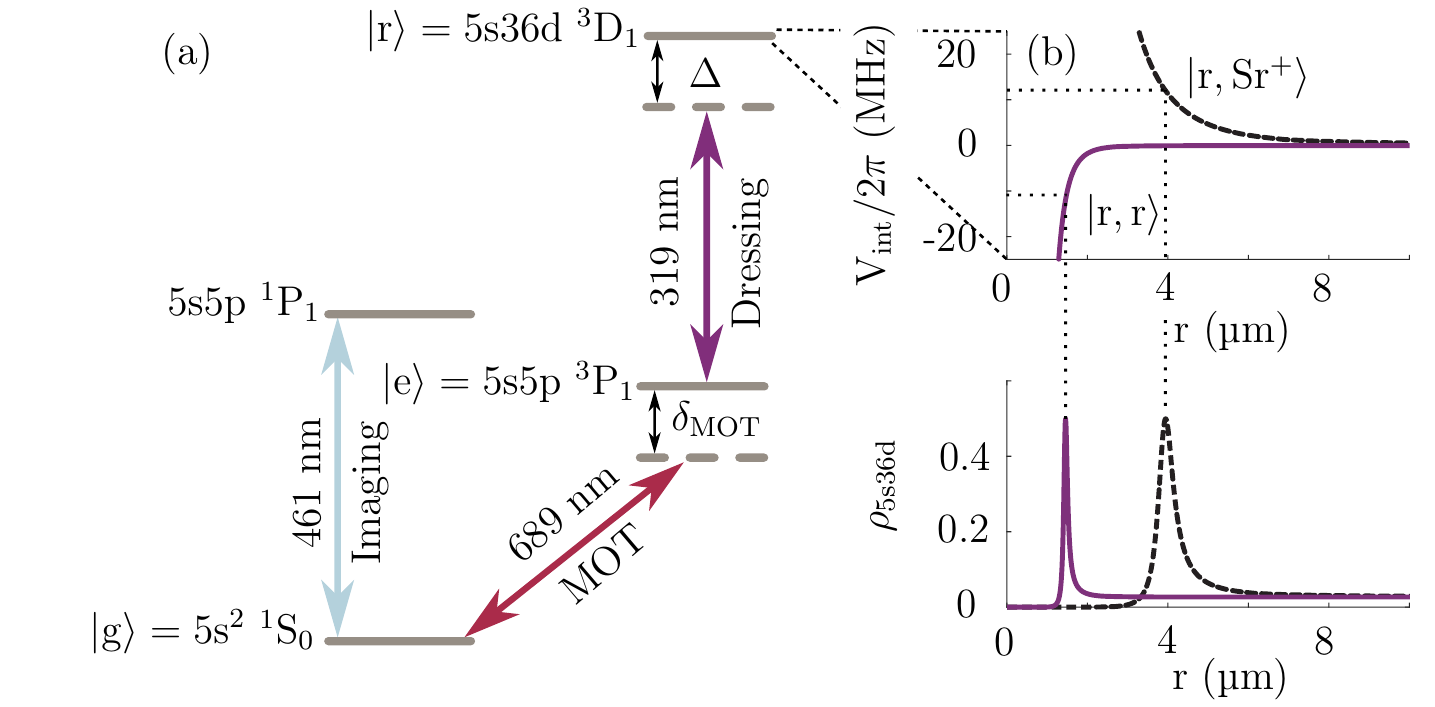}
\caption[Concept]{\label{fig:fig1} (a) Relevant energy levels in $^{88}$Sr. The MOT operates on the 689~nm transition and dressing using 319~nm light off-resonantly couples the $\ket{\mathrm{e}} = \mathrm{5s5p \ ^{3}P_{1}}$ state to the $\ket{\mathrm{r}} = \mathrm{5s36d \ ^{3}D_{1}}$ state. (b) Upper panel: atom-atom (solid purple line) and atom-ion (dashed black line) interaction strength as a function of interparticle separation $r$. Lower panel: the corresponding distance-dependent Rydberg excitation fraction for a detuning $\Delta =- (+) 12$~MHz for the atom-atom (atom-ion) pair. }
\end{figure}

	Recently our group successfully Rydberg dressed a magneto-optical trap (MOT) using the energy level scheme illustrated in figure \ref{fig:fig1}(a), achieving a lifetime of $>1$~ms \cite{Bounds2018}. During the course of this work, we observed that Rydberg dressing in large samples may be accompanied by the formation of an ultra-cold plasma (UCP), leading to rapid loss. While previous work has studied the coupling of a Rydberg gas to an UCP in both the pulsed \cite{Vitrant1982, Robinson2000,Millen2010} and continuous \cite{PhysRevLett.110.045004,PhysRevA.86.020702} regimes, these studies have focussed on the case of resonant excitation. In this paper, we study the growth of a plasma under off-resonant excitation using time-resolved detection of individual ions. We show that plasma formation occurs via a  Coulomb anti-blockade mechanism, illustrated in figure \ref{fig:fig1}(b), where background ions DC Stark shift nearby atoms into resonance at specific distances. The result is a highly correlated `facilitated growth' of the Rydberg population, that shares some similarities with that previously observed using van der Waals anti-blockade \cite{Amthor2010}. A rate equation model that couples the driven Rydberg gas to an UCP via this correlated excitation model accurately reproduces both the initial creation and subsequent decay of the observed ion signal. At longer times, the ionization of long-lived high angular momentum Rydberg states produced in the plasma provides evidence for the interplay between Coulomb blockade \cite{Engel2018} and Coulomb anti-blockade in the plasma.  In future experiments, controlling this process may provide a route to creating an UCP with a strongly correlated ionic state \cite{Bannasch2013}. Our work also provides a strategy for minimising losses due to ionization in cold Rydberg gas experiments.

\section{Experimental techniques}

The experiments began with the formation of a MOT of $^{88}$Sr atoms operating on the narrow $5\mathrm{s}^2 \ ^{1}\mathrm{S}_{0} \leftrightarrow 5\mathrm{s} 5\mathrm{p} \ ^{3}\mathrm{P}_{1}$ transition. A MOT beam detuning of $\delta_{\mathrm{MOT}} / 2 \pi \approx -175$~kHz and an intensity per beam of  $\mathrm{I_{MOT}} =3$-$50 \ \mathrm{I_{sat}}$ (where $\mathrm{I_{sat}}$ is the saturation intensity of the transition) resulted in  $1/e^{2}$ cloud radii of 30-60 \SI{}{\micro \metre} (100-200 \SI{}{\micro \metre}) in the vertical ($z$) (horizontal ($x$)) direction, and temperature $T_\mathrm{z}= $ \SI{1}{\micro K}. The MOT is loaded from a first-stage  MOT operating on the much broader  $5\mathrm{s}^2 \ ^{1}\mathrm{S}_{0} \leftrightarrow 5\mathrm{s}5\mathrm{p} \ ^{1}\mathrm{P}_{1}$ transition. By varying the number of atoms captured in the first-stage MOT, the final density of the MOT can be varied independently of the other experimental parameters. More details on the apparatus and loading process can be found in 
 \cite{BoddyThesis,SadlerThesis,BoundsThesis}.

Atoms in the MOT were subsequently coupled to the Rydberg state by off-resonantly driving the  
 $ 5\mathrm{s} 5\mathrm{p} \ ^{3}\mathrm{P}_{1} \rightarrow 5\mathrm{s}36\mathrm{d} \ ^{3}\mathrm{D}_{1}$ transition using  50~mW of 319~nm light \cite{Bridge2016} with a beam waist of \SI{160}{\micro \metre} (\SI{120}{\micro \metre}) in the vertical (horizontal) direction, resulting in a peak coupling beam Rabi frequency of $\Omega / 2\pi = 4$~MHz. The coupling beam detuning was varied from $\Delta / 2 \pi= -20$ to $+20$~MHz, allowing observation of the attractive van der Waals interaction of the $36\mathrm{d} \ ^{3}\mathrm{D}_{1}$ state for negative detunings, and the DC Stark shift due to nearby ions for positive detunings, illustrated in figure \ref{fig:fig1}(b).  As we showed in our previous work, laser cooling and trapping forces continue to operate in the Rydberg-dressed MOT \cite{Bounds2018}.
After dressing for a time $t_\mathrm{d}$, the cloud was imaged using resonant absorption imaging on the 461~nm transition (figure \ref{fig:fig1}(a)). As the cloud was often optically thick, ballistic expansion was sometimes used to reduce the optical depth of the cloud $\beta$ before imaging, sacrificing information on the cloud spatial profile. Fitting to the cloud shape, the ground state population $N_{\mathrm{g}}$ was measured.

The spontaneous generation of ions in the cloud \cite{Robinson2000,1367-2630-11-1-013052,PhysRevA.76.054702} was studied by continuously counting the ions using a micro-channel plate (MCP) rather than directly (and destructively) probing the Rydberg population through field- or auto-ionization. Detection was enhanced through a small bias field of $\sim$20~ mV/cm that draws ions towards the MCP in \SI{30}{\micro s}. Using a fast digital oscilloscope, the arrival of individual ions was logged with a detection of efficiency of up to 2\% \cite{BoundsThesis} and a time-resolution better than 10~ns \cite{Lochead2013}. The resulting ion signals were then binned according to their arrival times to produce an ion detection rate; by varying the bin width  rates could be measured over several orders of magnitude. The combination of non-destructive and time-resolved single-ion detection allows observation of ion detection rates in real-time over the long ($> 5$~ms) timescales relevant to Rydberg-dressing experiments, illustrated in figure \ref{fig:fig4}.
By repeating the experiment up to 12 times for a given set of experimental conditions, we gain access to the statistical distribution of the ion signal. In particular, we can evaluate the time-dependent Mandel Q parameter given by \cite{Mandel:79} 
\begin{equation} Q = \frac{\langle ( \Delta N(t))^{2}\rangle}{\langle N(t) \rangle} - 1 \ ; \end{equation}
where $\langle ( \Delta N(t))^{2}\rangle$ and $\langle N(t) \rangle$ are respectively the variance and mean of the ion signal at time $t$. Q parameters greater than zero indicate super-Poissonian signals often associated with anti-blockade and collective excitation; Q parameters below zero indicate sub-Poissonian statistics often associated with blockade effects and nonlinear excitation suppression \cite{PhysRevLett.109.053002}. We do not correct measured Q parameters for detection efficiency.

\section{Atom loss and spontaneous ionization}
\begin{figure}[h]
\centering
\includegraphics[width=13.3cm]{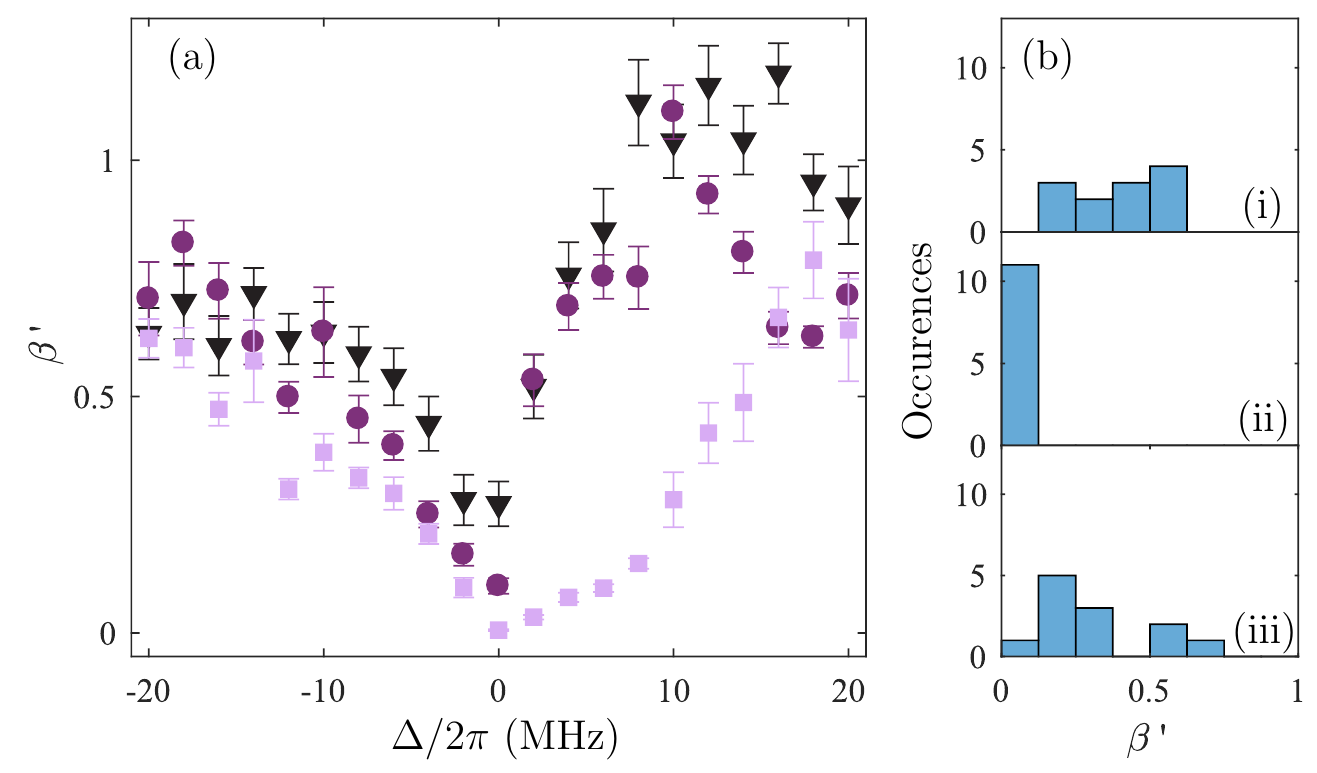}
\caption[Atom loss in a Rydberg dressed MOT]{\label{fig:fig2} (a) The fractional change $\beta^\prime$ = $\beta(t=t_\mathrm{d})/\beta(t=0)$ in optical depth $\beta$ after dressing for a time $t_\mathrm{d}= 5$~ms versus detuning for $\rho_{0} = 0.6 \times 10^{10}$~cm$^{-3}$ (black triangles), $2 \times 10^{10}$~cm$^{-3}$ (purple circles) and  $7 \times 10^{10}$~cm$^{-3}$ (pink squares). Here  $\delta_{\mathrm{MOT}}/2 \pi = -210$~kHz, $\mathrm{I_{MOT}} = 50 \ \mathrm{I_{sat}}$ and $\Omega / 2 \pi = 4$~MHz.  (b) Histograms showing the distribution of $\beta^\prime$ over multiple runs of the experiment for the high density data in (a)  (pink squares) for three different values of the detuning $\Delta = -10$~MHz (i), $0$~MHz (ii) and $+10$~MHz (iii).}
\end{figure}

The change in the optical depth of the MOT due to the Rydberg coupling laser is illustrated in figure \ref{fig:fig2}, for three different values of the initial density $\rho_0$. The lineshape of the loss feature is asymmetic. For the lowest and intermediate density, additional loss occurs for negative detunings relative to positive detunings. 

This result is consistent with the attractive van der Waals interaction \cite{Vaillant2012} shifting nearby pairs of atoms into resonance (anti-blockade) as observed in previous experiments \cite{PhysRevLett.98.023002,Amthor2010}.  A   density of $2 \times 10^{10}$~cm$^{-3}$ corresponds to a Wigner-Seitz radius (defined as the radius of a sphere of volume equal to the mean volume per atom) of \SI{2.3}{\micro \metre}, similar to the \SI{1.6}{\micro \metre} facilitation radius of the $\rm{36d \ ^{3}D_{1}}$ Rydberg state for $\Omega / 2 \pi = 4$~MHz, $\Delta / 2 \pi = -6$~MHz at which the van der Waals interaction shift is equal to the coupling beam detuning.

As the density was increased further, additional loss also appeared for positive detunings. Unlike for zero and negative detunings, the atom number was observed to fluctuate dramatically from shot to shot  as shown in figure \ref{fig:fig2}(b). This behaviour cannot be explained through the attractive van der Waals interactions of the $\rm{5s36d \ ^{3}D_{1}}$ Rydberg state. Instead we attribute this loss to the presence of ions in the cloud; for positive coupling beam detunings the ions DC Stark shift atoms into resonance with the coupling beam, leading to enhanced Rydberg excitation and atom loss from the MOT.

\begin{figure}[h]
\centering
\includegraphics[width=16cm]{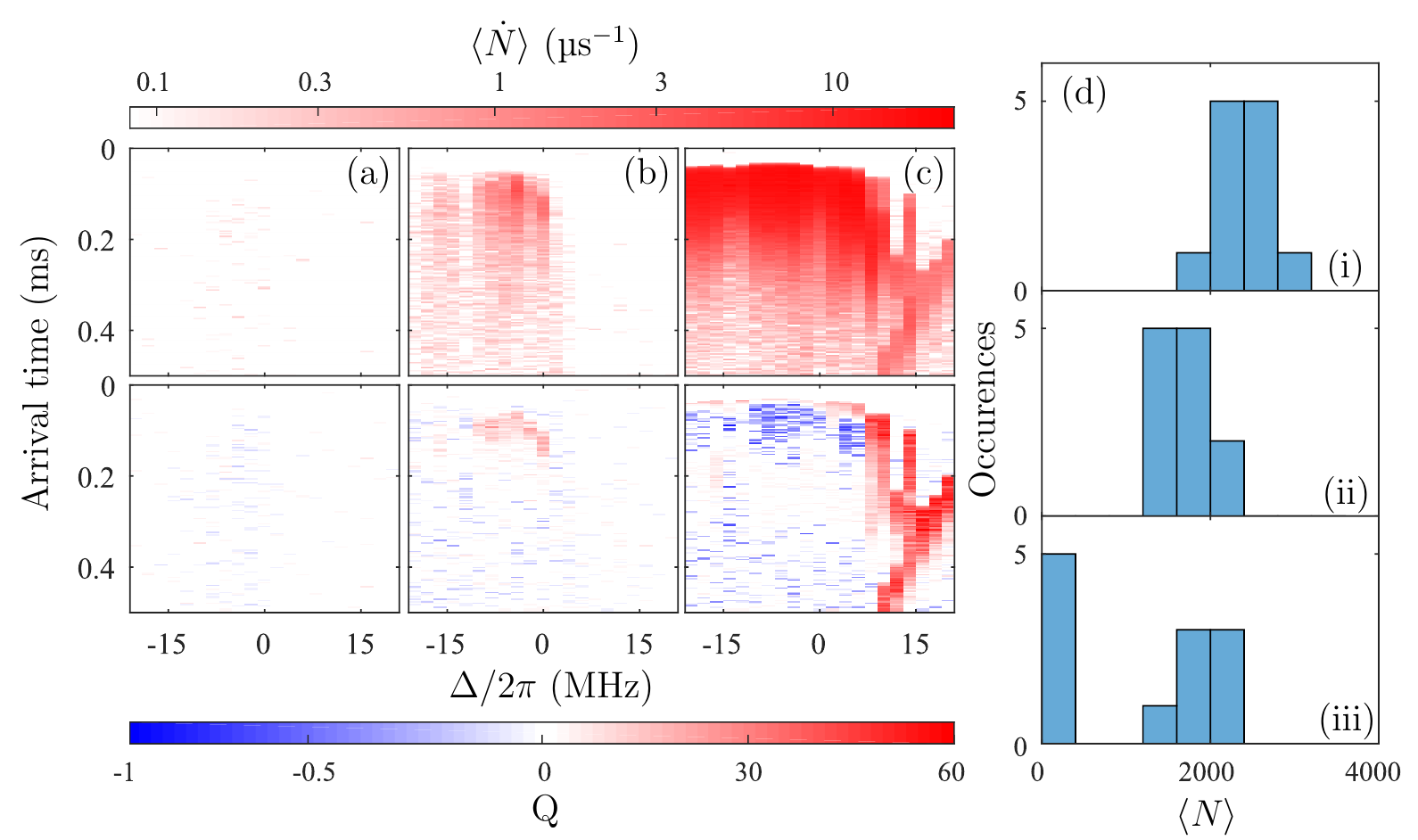}
\caption[Ion signal and loss in a Rydberg dressed MOT]{\label{fig:fig3} (a--c) Ion detection rate (top row) and Mandel Q parameter (bottom row) versus ion arrival time (measured from the start of the dressing pulse $t=0$) and detuning $\Delta$, for $\rho_{0}$ of (a): $0.6 \times 10^{10}$~cm$^{-3}$, (b) $2 \times 10^{10}$~cm$^{-3}$ (c): $7 \times 10^{10}$~cm$^{-3}$. (d) Histograms of the total detected signal  for all arrival times $\langle N\rangle$ for three different  values of the detuning $\Delta= -10$~MHz (i), $0$~MHz (ii) and $+10$~MHz (iii), for the highest density data shown in (c).  }
\end{figure}

The corresponding ion signals for the data shown in figure \ref{fig:fig2} are shown in figure \ref{fig:fig3}. The time dependence of both the ion detection rate $\langle \dot{N} \rangle$ and the Mandel Q parameter are shown. The ion detection rate shows a strong density dependence, with negligible ionization at low density. At intermediate densities, ions are only observed on resonance and for negative detunings. This observation is consistent with the enhanced loss for negative detunings observed in figure \ref{fig:fig2}. The ion signal is slightly super-Poissonian, which is consistent with the expected correlated growth of Rydberg excitations due to the anti-blockade effect \cite{PhysRevLett.109.053002}. This effect has been studied in detail elsewhere \cite{Malossi2012,Urvoy2015,Simonelli2016} and we do not examine these results further.

At the highest atom density (figure \ref{fig:fig3}(c)) significant ionization is also observed for negative detunings. In fact for $\Delta <0$ the detector saturates at high density, leading to the sub-Poissonian statistics observed in the Mandel Q parameter. Of interest for  this paper is the appearance of a large ion signal for positive detunings, accompanied by super-Poissonian statistics, which is consistent with the enhanced atom loss observed in figure \ref{fig:fig2}. Furthermore, histograms of the ion signal (figure \ref{fig:fig2}) show that the distribution of the total number of ions detected becomes bimodal at high density and positive detuning. This observation is consistent with an avalanche effect that strongly enhances Rydberg excitation and ionization. Our interpretation is that the correlated growth of Rydberg excitation due to the DC Stark shifts from nearby ions eventually leads to the runaway formation of an UCP, leading to rapid ionization of nearly all the remaining Rydberg atoms.

\section{Time dependence of the ion signal}

\begin{figure}[h]
\centering
\includegraphics[width=15.5cm]{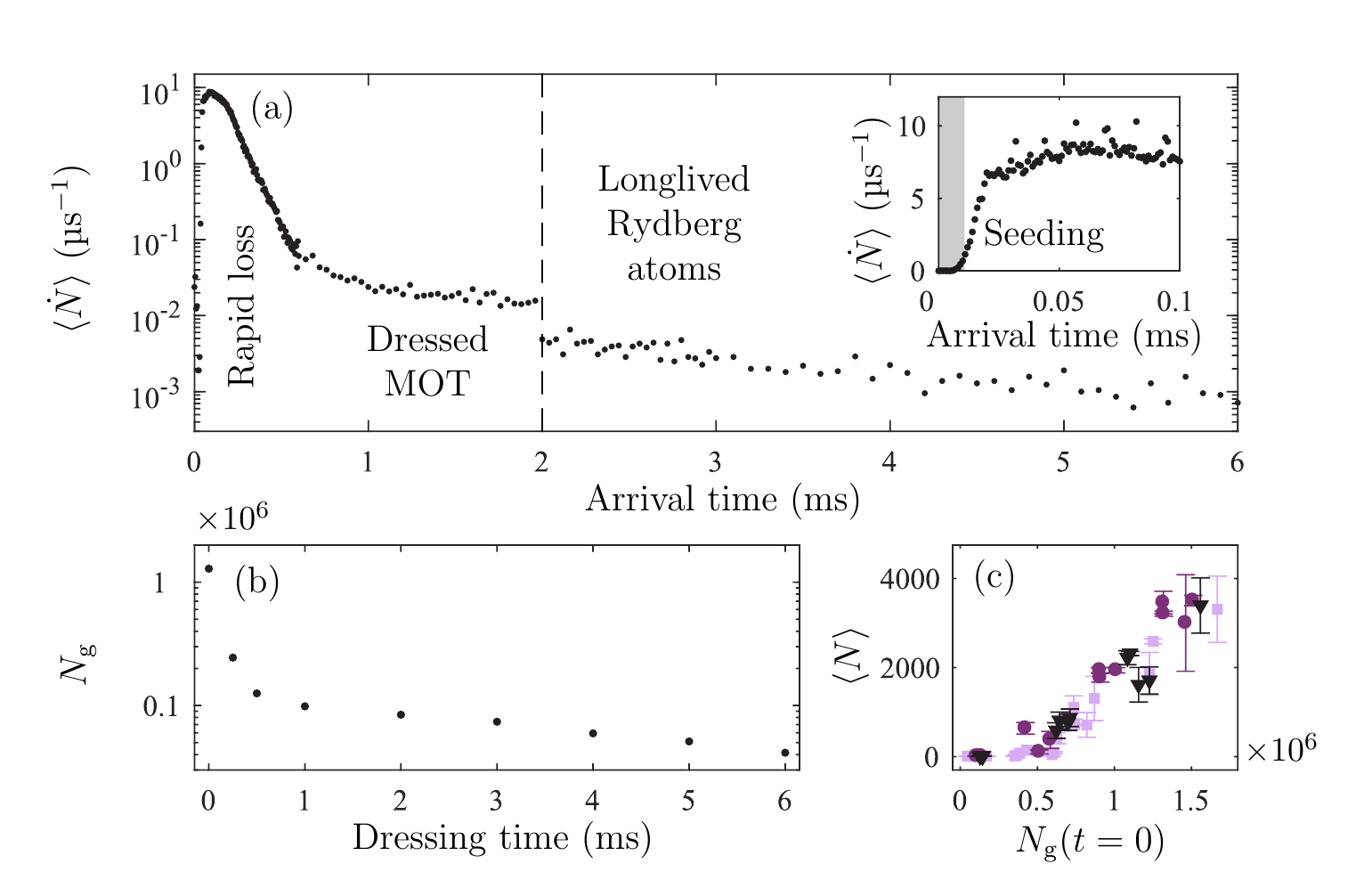}
\caption[Plasma threshold stages]{\label{fig:fig4} (a) Ion detection rate versus arrival time for an initial density $6 \times 10^{11}$~cm$^{-3}$ and $\Omega / 2 \pi = 4$~MHz, $\Delta / 2 \pi = +12$~MHz and a dressing time $t_\mathrm{d}$ =2 ms (indicated by the dashed line). The inset shows the first 0.1~ms of the same dataset with the plasma seeding time indicated by the shaded area. The \SI{30}{\micro s} for ions to reach the detector has been subtracted. (b) Measured variation in atom number versus $t_\mathrm{d}$. Error bars are too small to be seen. The point at $t_\mathrm{d}=2$ ms corresponds to the data shown in (a). (c) Variation of the total detected ion signal $\langle N \rangle$ with initial atom number for different values of the total power in the MOT beams within the ranges $<60$~\SI{}{\micro W} (pink squares), $60-150$~\SI{}{\micro W} (purple circles) and $>150$~\SI{}{\micro W} (black triangles).}
\end{figure}

To investigate this effect in more detail, we study the time dependence of the ionization rate for a single detuning. These data were taken with different parameters to figure \ref{fig:fig2} that are chosen such that the plasma threshold is always reached and the distribution is no longer bimodal. Four different stages can be identified in figure \ref{fig:fig4}:

\begin{itemize}
\item Plasma seeding and growth (inset (a)) - after a density-dependent delay (the plasma seeding time), the ion signal rapidly rises within the first \SI{20}{\micro s} of dressing. 

\item Rapid atom loss ((a) and (b)) - strong Rydberg excitation results in large ion signals and rapid loss of ground state atoms. The ion signal and the atom number decay at similar rates.

\item Dressed MOT -  the rate of atom loss reduces abruptly and the ion detection rate stabilises, leaving a relatively long-lived, low density Rydberg-dressed MOT that is similar to that presented in \cite{Bounds2018}.

\item Long-lived Rydberg atoms -  a long-lived  ion signal continues to be observed even when the dressing laser and MOT beams are switched off.

\end{itemize}

To interpret these stages, we note that previous work has shown that in a spontaneously ionizing Rydberg gas, an UCP is formed once the number of ions exceeds a threshold given by  $N_{C} = E_{e} \sqrt{(\pi / 2 )} \ 4 \pi \epsilon_{0}\sigma / e^{2}$, where the electron energy  $E_{e}$ is equal to the Rydberg binding energy, $e$ is the electron charge, $\epsilon_{0}$ is the electric permittivity of vacuum and $\sigma$ is the cloud radius \cite{Killian1999,Robinson2000,Li2004}.

The resulting plasma undergoes rapid expansion driven by disorder-induced heating \cite{Murillo2001,Simien2004}, typically living for just a few tens of microseconds \cite{Kulin2000}. During the plasma phase, the trapped electrons collide with the remaining Rydberg atoms, leading to rapid ionization and the formation of long-lived high-angular momentum Rydberg states \cite{Dutta2001}. 

The density-dependent seeding time thus represents the time taken for the number of ions to reach the plasma threshold. At this point, electron Rydberg collisions become possible, leading to a rapid increase in the ionization rate. The ions in the plasma are replenished from the (decaying) atomic population, until the rate of ionization is no longer high enough to balance the rate of ion loss due to plasma expansion, and the plasma is quenched. Ionization can still proceed through non-collisional processes, but at a much lower rate, and the atom population decay is reduced. Once the dressing laser is switched off, the slow black-body radiation induced ionization of long-lived, high-$l$ Rydberg atoms \cite{Dutta2001} created within the plasma is observed. In figure \ref{fig:fig4}(c) we show that the total number of detected ions exhibits the threshold dependence on the ground state atom number that is characteristic of an UCP.

\section{Theoretical description}
The coupled evolution of a Rydberg gas and an UCP has been considered theoretically by several groups \cite{PhysRevA.86.020702,PhysRevLett.110.045004,PhysRevA.89.022701}. However these works have focussed primarily on the role of ions in suppressing, rather than enhancing, the Rydberg excitation rate by DC Stark shifting the resonant Rydberg coupling laser off-resonance with the transition. Here we develop a model based on coupled rate equations for the total number of each of the three key species present in the cloud: ground state atoms $N_{\mathrm{g}}$, Rydberg atoms $N_{\mathrm{Ryd}}$ and ions in the cloud $N_{\mathrm{ion}}$. Solutions of the rate equations can then be compared directly to data such as that shown in figure \ref{fig:fig4}. 

The initial atom cloud is assumed to be a spheroid of randomly-distributed ground-state atoms, with uniform density and radius $\sigma$. As a further simplifying assumption, the Rabi frequency of the coupling laser $\Omega$ is also considered to be uniform across the cloud. A fraction $ P_{E}$ of the atoms are considered to be excited to the $\rm{5s5p \ ^{3}P_{1}}$ state by the cooling light in the MOT. The value of  $P_{E}$ depends on the MOT beam power, and is constrained within the range  $0.04 < P_{E} < 0.5$ by the operation of the MOT \cite{Loftus2004}.

\begin{figure}[h]
\centering
\includegraphics[width=15.72cm]{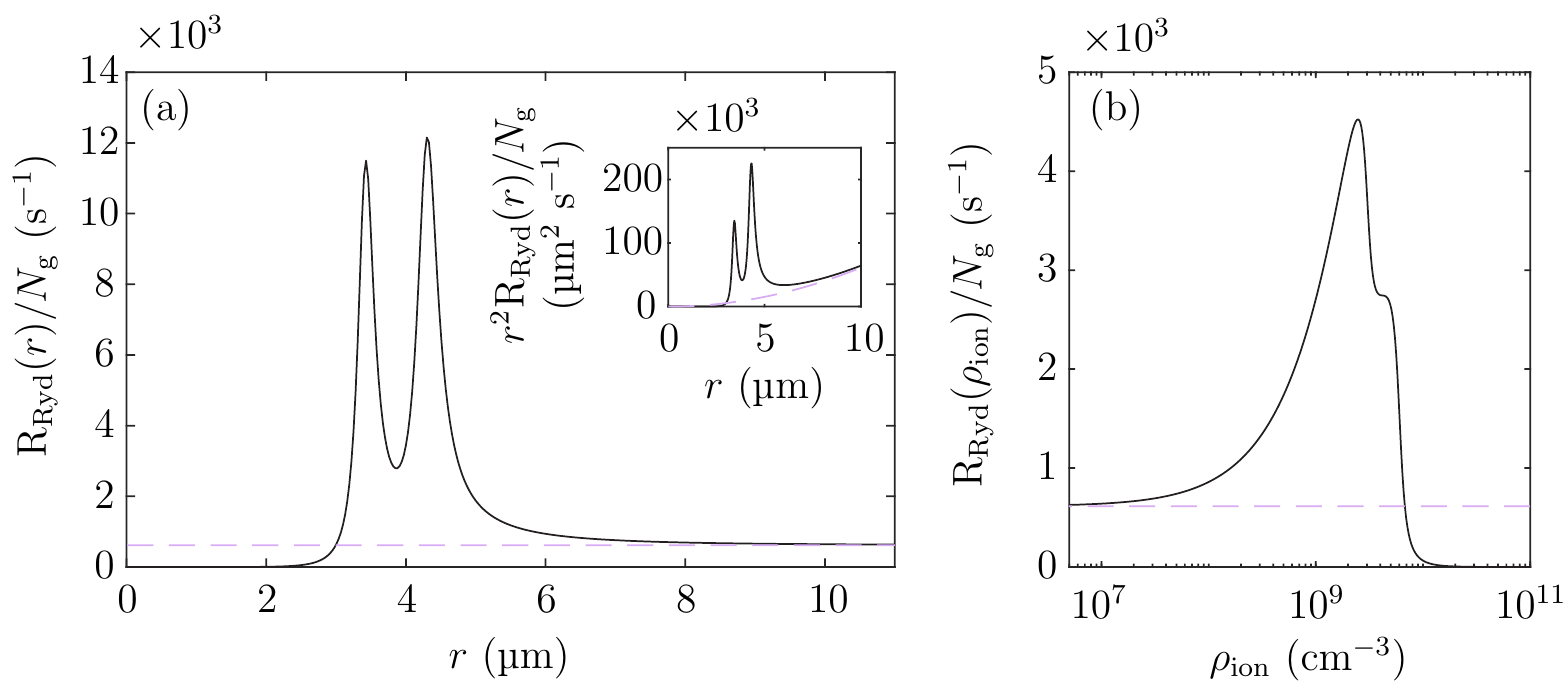}
\caption[Plasma model]{\label{fig:fig5} (a) Rydberg excitation rate as a function of distance from an ion, described by equation \ref{eq:summed:excitation_rate} for $P_{E} = 0.5$ (black solid line). The one-body rate is indicated by the pink dashed line.  Inset shows the excitation rate multiplied by $r^{2}$ for the same parameters, reflecting the integral weighting in equation \ref{eq:ion_anti_blockade}. (b) Integrated excitation rate as a function of ion density  $\rho_{\mathrm{ion}}$ (black solid line). The pink dashed line indicates the one-body excitation rate. }
\end{figure}

Atoms in the $\rm{5s5p \ ^{3}P_{1}}$ state are available for excitation to the Rydberg state by the coupling laser. As the timescale to reach the steady state is on the order of the Rydberg state lifetime of $1/{\Gamma_{\mathrm{Ryd}}} \sim$\SI{20}{\micro s}, it is much shorter than the timescale associated with the atom loss and ionization shown in figure \ref{fig:fig4}, allowing us to consider the steady-state Rydberg excitation rate described by the power-broadened ($\Omega \gg \Gamma_{\mathrm{Ryd}}$) Lorentzian lineshape:
\begin{equation} \label{eq:charge_enhanced_excitation}  R_{\mathrm{Ryd, |m_{J}|}}(r) = N_{\mathrm{g}} P_{\mathrm{E}} \frac{\Gamma_{\mathrm{Ryd}}}{2} \frac{\Omega^{2}}{\Omega^{2} + 2 [\Delta - \Delta_{\mathrm{DC, |m_{J}|}}(r)]^{2}}  \ . \end{equation}
Here the laser detuning $\Delta$ is combined with the DC Stark shift of the Rydberg state caused by the presence of an ion at a distance $r$
\begin{equation} \label{eq:dc_stark_shifts} \Delta_{\mathrm{DC}}(r) =  - \frac{1}{2} \alpha_{|m_{J}|} \Bigg[ \frac{1}{4 \pi \epsilon_{0}} \frac{e}{r^{2}} \Bigg]^{2} \ ; \end{equation}
 where the polarisability of the Rydberg state is $\alpha_{|m_{J}| = 0(1)} = - 2 \pi \times 40~(16)$~MHz/(V/cm)$^{2}$ for the two $\mathrm{36d \ ^{3}D_{1}} \ |m_{J}|$ states \cite{BoundsThesis}. To take into account the different static polarizability of the $|m_{J}|=0$ and $|m_{J}|=1$ sublevels we calculate the total excitation rate 
\begin{equation} \label{eq:summed:excitation_rate}  R_{\mathrm{Ryd}}(r) = R_{\mathrm{Ryd}, |m_{J}|=0}(r)  + R_{\mathrm{Ryd}, |m_{J}|=1}(r) \ ;  \end{equation}
assuming that the coupling strength for each is $\Omega/\sqrt{2}$ to reflect that on average only half of the coupling light will be the correct polarisation to excite each sublevel.
This term plays a key role in the simulation by allowing for the enhanced growth of Rydberg excitations when $\Delta \approx \Delta_\mathrm{DC}$, as shown in figure \ref{fig:fig5}. To include this effect in the model, we express the Rydberg excitation rate as a function of ion number $N_\mathrm{ion}$ rather than distance $r$, by integrating over the distribution of atom-ion separations. We treat ions as uniformly distributed such that the number of atoms around an ion is given by $\rho_{\mathrm{atom}} 4 \pi r^{2} \delta r$. The Rydberg excitation rate is then calculated as a function of ion number: 

\begin{equation} \label{eq:ion_anti_blockade} R_{\mathrm{Ryd}}(N_{\mathrm{ion}}) = \frac{\int_{0}^{r_{\mathrm{ion}}} R_{\mathrm{Ryd}}(r) \times \rho_{\mathrm{atom}} 4 \pi r^{2} \delta r}{\int_{0}^{r_{\mathrm{ion}}} \rho_{\mathrm{atom}} 4 \pi r^{2} \delta r}  \ . \end{equation}

The upper limit of this integral is set by the mean distance between ions given by the  Wigner Seitz radius $ r_{\mathrm{ion}} = ( 3 / 4 \pi \rho_{\mathrm{ion}} )^{1/3} =\sigma / N_\mathrm{ion}^{1/3}$, where $\rho_{\mathrm{ion}} = 3 N_{\mathrm{ion}} / 4 \pi \sigma^{3}$ is the ion density. The ion Wigner Seitz radius is the radius of a sphere with a volume equal to the mean volume per ion; using this approach is therefore equivalent to considering the cloud as comprised of spheres of atoms surrounding each ion. 

The way in which this term leads to density-dependent coupling between Rydberg atoms and ions, along with strong spatial correlations, is shown in figure \ref{fig:fig5}. At low charge densities, corresponding to large typical atom-ion separations, the Rydberg excitation rate follows the one-body rate. At higher charge densities the Rydberg excitation rate rises until at the `facilitation radius' the detuning $\Delta$ is compensated by the DC Stark shift. In figure \ref{fig:fig5} two peaks appear due to the presence of resonances associated with the $|m_{J}|=0$ and $|m_{J}|=1$ states. The result is a substantial increase in Rydberg excitation and results in the rapid loss of atoms observed in figure \ref{fig:fig4}. At ion densities above $2.5 \times 10^{9}$~cm$^{-3}$ this effect saturates and the integrated excitation rate begins to fall with density. Atoms are so close to ions that the DC Stark shift is large enough to shift atoms beyond resonance with the coupling laser. For large enough ion density, the integrated excitation rate falls below the one-body rate: this is the Coulomb blockade regime \cite{Engel2018}. This suggests that Coulomb blockade may result in a self-limiting Rydberg population similar to that proposed in \cite{PhysRevA.86.020702}, even when the overall effect is one of avalanche anti-blockade.

To complete the rate equations, we also consider the decay and loss mechanisms for Rydberg atoms and ions.  Our experiments have shown that there are two important one-body decay processes that contribute to the overall decay rate of the Rydberg states $\Gamma_\mathrm{Ryd}$. The first is radiative decay, which occurs predominantly to low-lying triplet states that are not connected by dipole-allowed transitions to the ground state. This decay is included in the model by introducing a dark state with  associated population $N_{\mathrm{dark}}$. Population in this state does not contribute to absorption images, and appears as an overall loss of atoms from the cloud. The second process is spontaneous ionization. Even in the absence of an UCP (i.e. below threshold), a significant background ionization rate is observed.  To quantify the branching ratio between radiative and ionizing decay  we have measured the ratio of ions created without autoionization (i.e. through spontaneous ionization) to the total Rydberg population measured using autoionization \cite{SadlerThesis} following a controlled excitation sequence with no plasma or significant ion population present. From this we estimate that approximately $P_{\mathrm{ion}} = 30$\% of the Rydberg atoms decay to ions, through several mechanisms such as blackbody radiation \cite{1367-2630-11-1-013052}, collision with hot atoms \cite{Robinson2000} and Penning ionization \cite{PhysRevA.76.054702}. A high ionization fraction is consistent with high densities of attractively interacting Rydberg states \cite{PhysRevLett.100.123007,Viteau2008}. The spontaneous ionization process is included in the model as a one body coupling $\dot{N}_{\mathrm{ion}}= P_{\mathrm{ion}} \Gamma_\mathrm{Ryd} N_\mathrm{Ryd}$ with decay to dark states described as $\dot{N}_{\mathrm{dark}}= (1 - P_{\mathrm{ion}}) \Gamma_\mathrm{Ryd} N_\mathrm{Ryd}$.

Crucially, above the threshold for plasma formation, Rydberg atoms are also ionized by collisions with  trapped electrons \cite{Vitrant1982}, resulting in an avalanche process as the rising electron population increases the collision rate.  In our simulations, we consider that a plasma forms when $N_{\mathrm{ion}} > N_{C} = E_{e} \sqrt{(\pi / 2 )} \  4 \pi \epsilon_{0}\sigma / e^{2}$, where the electron energy  $E_{e}$ is equal to the Rydberg binding energy \cite{Li2004}. Above this threshold, an additional term proportional to the electron density couples $N_\mathrm{Ryd}$ and $N_\mathrm{ion}$, given by $\dot{N}_\mathrm{ion}= 3 \gamma_{\mathrm{coll}} N_{\mathrm{Ryd}} N_{e}/ 4 \pi \sigma_{\mathrm{P}}^{3}$, where $N_{e} = N_{\mathrm{ion}}-N_{C}$ is the number of trapped electrons, and $\sigma_{\mathrm{P}}$ is the plasma radius, so that $3 N_{e}/ 4 \pi \sigma_{\mathrm{P}}^{3}$ is the electron density. The rate of ionizing electron-Rydberg collisions is given by $\gamma_{\mathrm{coll}} = \sigma_{\mathrm{geo}} \sqrt{E_{e} / m_{e}}$, where $\sigma_{\mathrm{geo}} = \pi n^{4} a_{0}^{2}$ is the geometric cross-section of the Rydberg atom for a principal quantum number of $n$ \cite{PhysRevLett.110.045004}. 

The final components of the model are terms that govern the rate at which ions are expelled from the cloud. We consider two processes. First, an applied electric field of $\sim$~20~mV/cm extracts ions from the cloud in $\sim$~\SI{7}{\micro s}. This extraction time gives rise to an ion loss rate of $R_{\mathrm{extract}}$.  These ions, labelled $N_{\mathrm{free}}$, are then too far from the cloud to play any further part in the system dynamics so are considered lost. The second process is loss due to expansion of the plasma. Previous experiments on UCPs have shown that the dominant driver of expansion is the back-action of the electrons on the ions (electron pressure) \cite{Kulin2000}, giving rise to a loss term of the form $\dot{N}_{\mathrm{free}} =  N_{\mathrm{ion}} N_{e} R_{\mathrm{repulsion}}$ where $R_{\mathrm{repulsion}}$ is treated as a free parameter.

Taken together, these relationships lead to the following rate equations:

\begin{equation} \label{eq:1} \dot{N}_{\mathrm{g}} = - R_{\mathrm{Ryd}}(N_{\mathrm{ion}}) \ ;  \end{equation}

\begin{equation} \label{eq:2} \dot{N}_{\mathrm{Ryd}} = + R_{\mathrm{Ryd}}(N_{\mathrm{ion}}) - \Gamma_{\mathrm{Ryd}} N_{\mathrm{Ryd}} -  \gamma_{\mathrm{coll}} N_{\mathrm{Ryd}} N_{e} /  \frac{4}{3} \pi \sigma_{\mathrm{P}}^{3} \ ;  \end{equation}

\begin{equation} \label{eq:3} \dot{N}_{\mathrm{ion}} = + P_{\mathrm{ion}} \Gamma_{\mathrm{Ryd}} N_{\mathrm{Ryd}} +  \gamma_{\mathrm{coll}} N_{\mathrm{Ryd}} N_{e} /  \frac{4}{3} \pi \sigma_{\mathrm{P}}^{3} - N_{\mathrm{ion}} R_{\mathrm{extract}} - N_{\mathrm{ion}} N_{e} R_{\mathrm{repulsion}} \ ; \end{equation}

\begin{equation} \label{eq:4} \dot{N}_{\mathrm{free}} = +N_{\mathrm{ion}} R_{\mathrm{extract}} + N_{\mathrm{ion}} N_{e} R_{\mathrm{repulsion}} \ ; \end{equation}

\begin{equation} \label{eq:5} \dot{N}_{\mathrm{dark}} = + (1 - P_{\mathrm{ion}}) \Gamma_{\mathrm{Ryd}} N_{\mathrm{Ryd}} \ . \end{equation}

If the plasma size $\sigma_{\mathrm{P}}$ is taken to be equal to the size of the atomic cloud, then only two free parameters are not fixed by measurement or theory: $P_\mathrm{E}$ (which is constrained) and $R_{\mathrm{repulsion}}$.  The ion detection efficiency (a linear scaling between the measured signal and $N_\mathrm{ion}$) is also treated as a free parameter, with best fitting found with a factor of 0.1\%. 

Before presenting a  detailed comparison of the model with the measured data  in the next section, it is insightful to reconsider some of the underlying physics. In particular, we point out that our model assumes that the formation of the plasma introduces an additional ionization channel, but does not modify the ion-dependent excitation process. In order for this to be the case, the length scale for the correlated excitation of Rydberg atoms due to nearby ions must be shorter than the Debye screening length $\lambda_{\mathrm{D}} = \sqrt{\epsilon_{\mathrm{0}} k_{\mathrm{B}} T_{\mathrm{e}} / e^{2} n_{e}}$ \cite{Killian1999}. The electron temperature is given by $T_{\mathrm{e}}$, and $n_{e}$ is the electron density. For typical values, we expect a Debye length greater than \SI{10}{\micro \metre}, much greater than the interparticle separations. Consequently, we do not expect electrons within the plasma to screen atoms in the cloud from the presence of nearby ions, and do not consider them to modify the steady-state Rydberg excitation rate.

\section{Comparison with experimental data}
\label{sec:lifetime}

\begin{figure}[h]
\centering
\includegraphics[width=15.32cm]{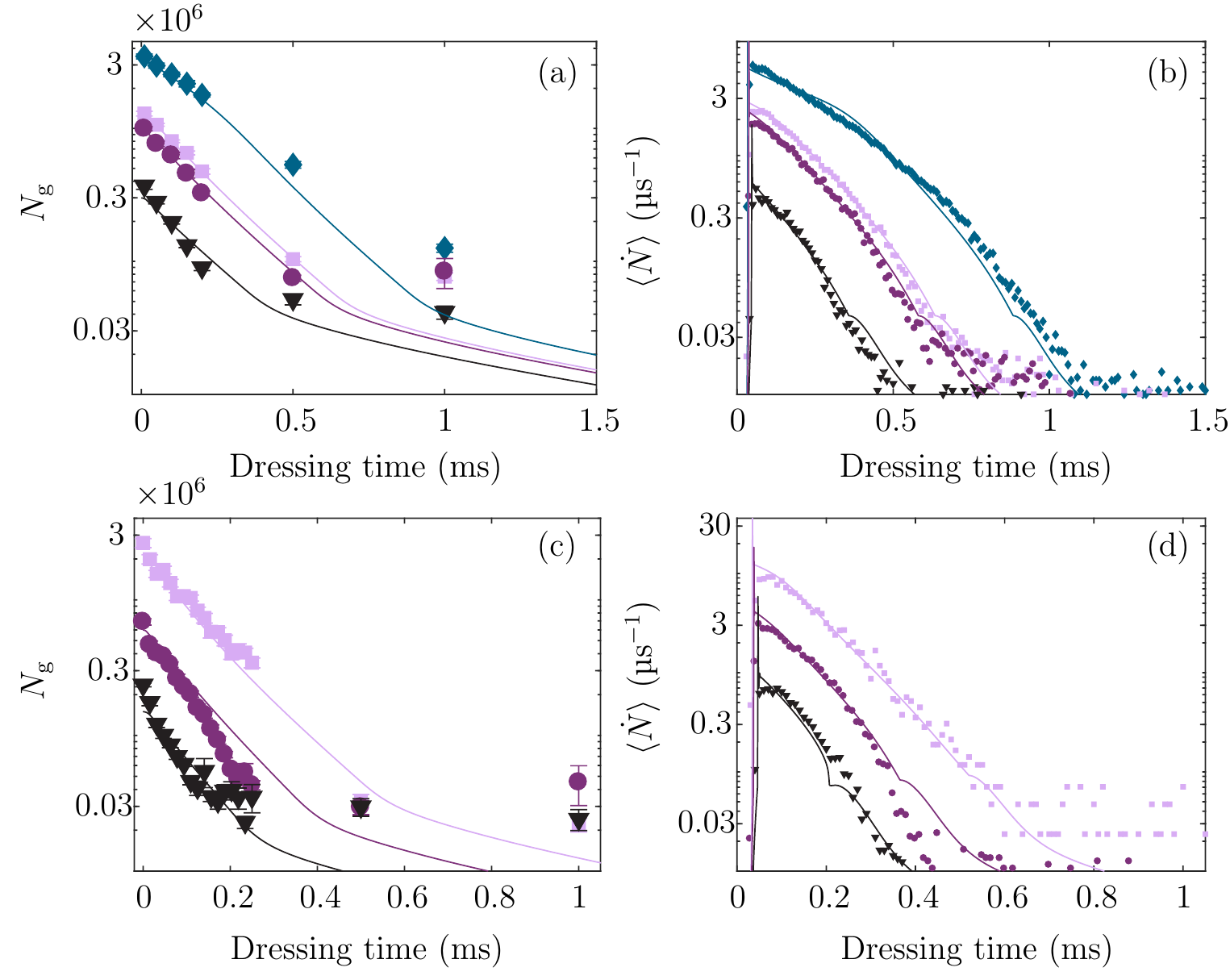}
\caption[Plasma lifetime]{\label{fig:fig6} Ground state atom number (a, c) and ion detection rate (b, d) as a function of dressing time for different initial atom numbers at $\delta_{\mathrm{MOT}} / 2 \pi = -110$~kHz, $\Delta / 2 \pi= +12$~MHz, $\Omega / 2 \pi = 4$~MHz and  MOT beam intensities of $\mathrm{I_{MOT}} = 9 \ \mathrm{I_{sat}}$ (a, b) and   $\mathrm{I_{MOT}} = 26 \ \mathrm{I_{sat}}$ (c, d). Model results are shown as solid lines and different colours and markers indicate the initial atom number, shown on the y-axis of (a) and (c). }
\end{figure}

In this section, we compare solutions of the rate equations (equations \ref{eq:1}-\ref{eq:5}) to experimental measurements of the atom number and ion signal. The best fit illustrated in figure \ref{fig:fig6} was obtained with  $R_{\mathrm{repulsion}} = 10^{4} e^{-1} s^{-1}$ and $P_{\mathrm{E}} = 0.5 \ (0.3)$ for a MOT beam intensity of  $\mathrm{I_{MOT}} = 26 \ \mathrm{I_{sat}}$ ($\mathrm{I_{MOT}} = 9 \ \mathrm{I_{sat}}$). This value of  $R_{\mathrm{repulsion}} = 10^{4} e^{-1} s^{-1}$  corresponds to ions leaving the cloud in \SI{1}{\micro s} for an electron population of 100, which is consistent with plasma expansion velocities observed in UCPs \cite{Kulin2000}. The plasma size $\sigma_{\mathrm{P}}$ was also treated as a free parameter. For the case of  $\mathrm{I_{MOT}} = 9 \ \mathrm{I_{sat}}$, we see good agreement with a constant plasma size of \SI{45}{\micro \metre}, consistent with the cloud size of \SI{50}{\micro \metre} by \SI{100}{\micro \metre}  $1 / e^{2}$ vertical and horizontal radii. However, in the case of higher MOT beam power, we observe best agreement allowing the plasma size to rise with initial atom number from \SI{40}{\micro \metre} to \SI{50}{\micro \metre}, suggesting that at sufficiently high ion production rates the plasma size is a function of ion production rate.

Figure \ref{fig:fig6} shows that the simulation accurately reproduces the temporal behaviour of the atom number and the ion signal over several orders of magnitude, at least for the first $\sim1$ ms of dressing. This includes both the rapid rise of the ion signal during the initial growth phase, and the subsequent decay, including regions where this decay deviates from a simple exponential. Importantly, examining each of the terms in this model enables us to understand which processes are important in each phase.

\begin{figure}[h]
\centering
\includegraphics[width=14.26cm]{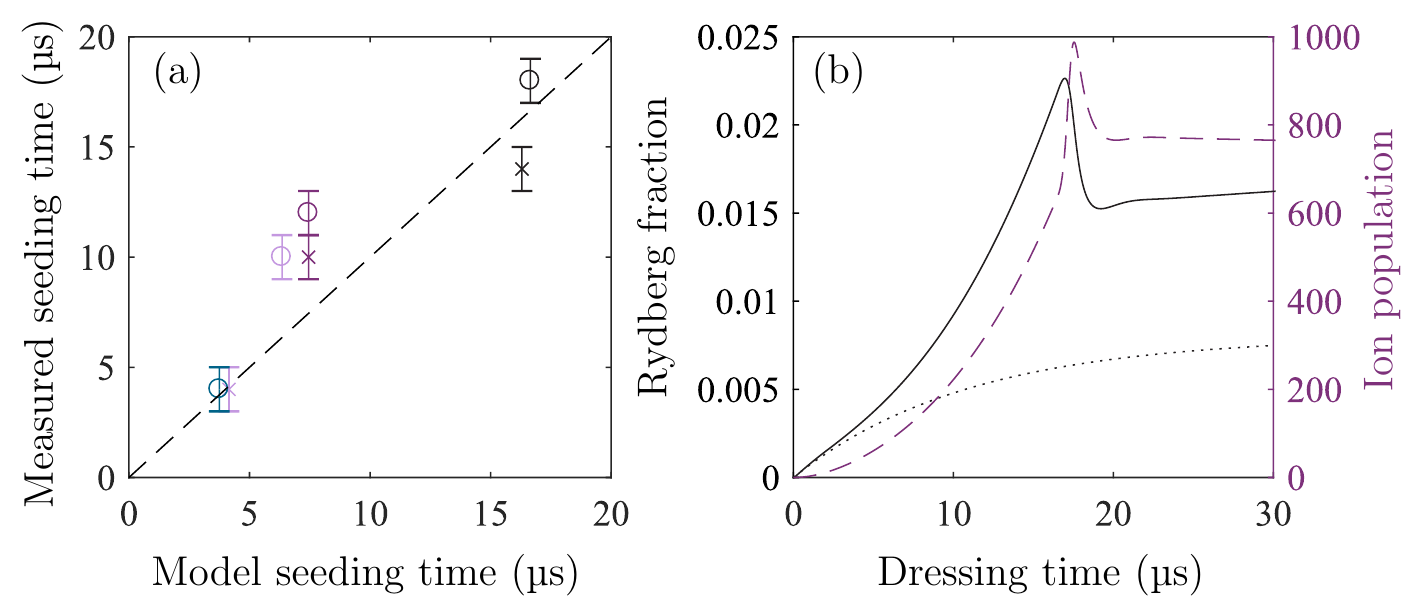}
\caption[Plasma seeding]{\label{fig:fig7} (a) Measured plasma seeding time against simulated plasma seeding time, for  $\mathrm{I_{MOT}} = 9 \ \mathrm{I_{sat}}$ (circles) and  $\mathrm{I_{MOT}} = 26 \ \mathrm{I_{sat}}$ (crosses). Colours indicate initial atom number and correspond to the colours used in figure \ref{fig:fig6}. The \SI{30}{\micro s} time for ions to reach the detector has been subtracted. (b) The simulated Rydberg fraction (black line) and the ion population (purple dashes) as a function of dressing time, simulated for  $\mathrm{I_{MOT}} = 9 \ \mathrm{I_{sat}}$ and an initial atom number of 0.3~million. The dotted black line shows the Rydberg fraction in the absense of charges.}
\end{figure}

The density-dependent plasma seeding time defined in figure \ref{fig:fig4} is also quantitatively reproduced by the model, as shown in figure \ref{fig:fig7}. The model shows that the seeding time and subsequent rapid loss depend critically on the existence of the Coulomb anti-blockade mechanism illustrated in figure \ref{fig:fig5}. As shown in figure \ref{fig:fig7}(b), the rise in the Rydberg population only occurs if the correlated growth of the Rydberg fraction due to ions is taken into account. The delay between the rise in Rydberg fraction and the rise in ion population is attributed to the time taken for ionization of Rydberg atoms to occur. When the plasma threshold is reached, the model predicts a sharp spike in the number of ions that dissipates due to plasma expansion that we do not observe, likely due to ions spreading out on their way to the detector.

\begin{figure}[h]
\centering
\includegraphics[width=15.19cm]{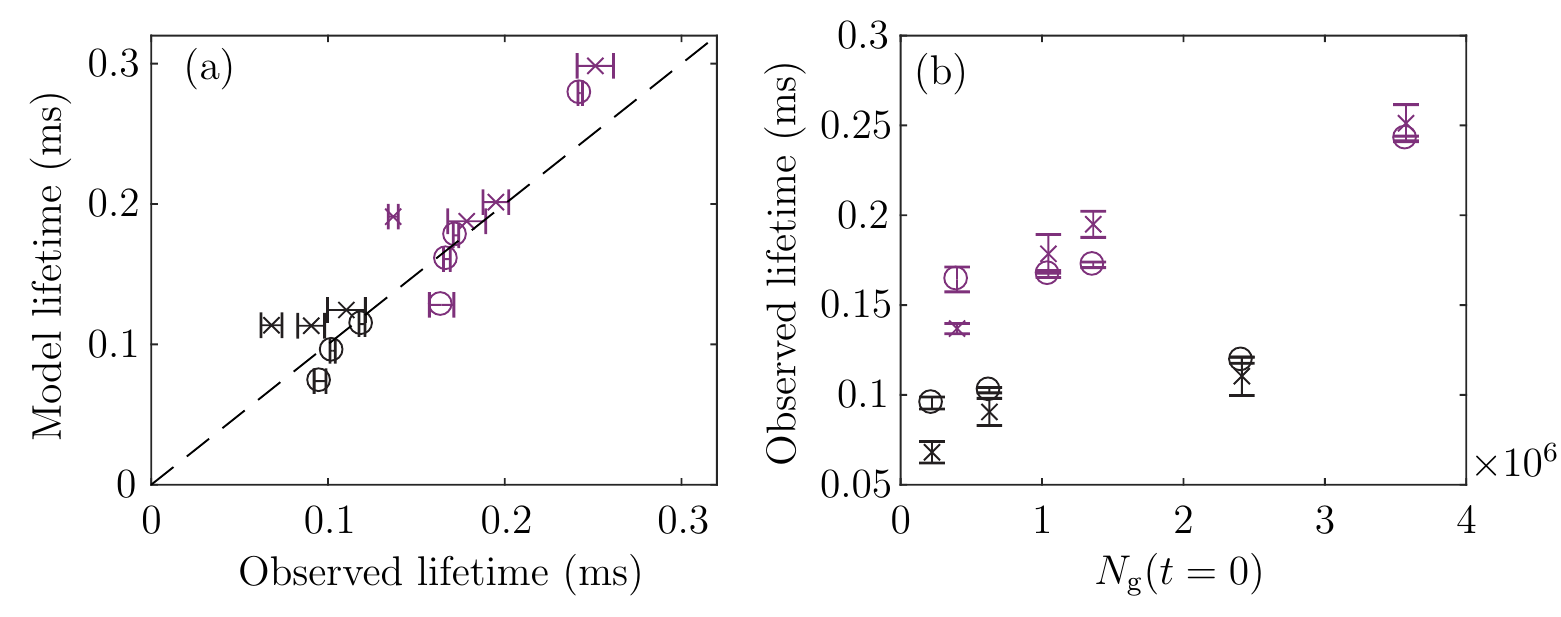}
\caption[Plasma lifetime]{\label{fig:fig8} (a) Simulated atom (crosses) and plasma (circles) lifetimes for MOT beam powers of   $\mathrm{I_{MOT}} = 26 \ \mathrm{I_{sat}}$ (black) and  $\mathrm{I_{MOT}} = 9 \ \mathrm{I_{sat}}$ (purple) as a function of observed lifetimes, from exponential fits to the data. (b) Observed atom and plasma lifetimes as a function of initial atom number.}
\end{figure}

Once the plasma has formed, collisional ionization of Rydberg atoms by electrons becomes a dominant process. The population in the Rydberg state drops as the ionization rate due to Rydberg-electron collisions dominates. The result is a largely exponential decay of the atom number, which since the ions are continually lost from the cloud is also mirrored in the ion signal. A quantitative comparison between the model and the data in this regime was made by fitting an exponential decay to the experimental and simulated data shown in  figure \ref{fig:fig6}. The resulting $1/e$ lifetimes are compared in figure \ref{fig:fig8}(a), and the overall agreement is very good. We also observe a rise in plasma and cloud lifetime with increasing initial atom number, shown in figure \ref{fig:fig8}(b). At first glance this result is suprising. The origin of this effect is the switch from Coulomb anti-blockade to Coulomb blockade for charge densities above $2.5 \times 10^{9}$~cm$^{-3}$ shown in figure \ref{fig:fig5}(b). As the atom number increases, the charge density increases above this value, leading to a reduction in the Rydberg fraction and an increase in the cloud and plasma lifetime. The plasma simulation predicts a peak charge density of up to $4 \times 10^{9}$ cm$^{-3}$ across the plasma, sufficient to reach the self-limiting regime. This self-limiting effect is also responsible for the deviation from exponential decay in atom number and ion detection rate observed in figure \ref{fig:fig6}, which is observed in both the data and the model. Further evidence of the important role of this density-driven cross-over from anti-blockade to blockade is presented in section 7.

Eventually, the number of ions in the cloud decays to below the plasma threshold, and the plasma terminates, resulting in the  abrupt change in the slope of the decay for atoms and ions that is observed in the atom number between 0.5 ms and 1 ms (figure~\ref{fig:fig6}). After this point, the role of ions in the excitation and decay processes becomes insignificant due to the low atom and ion density, and the rate equation model is no longer appropriate. Instead, the losses from the cloud are dominated by mechanical effects that originate from the combined effect of the AC Stark shift due to the dressing laser, and the ongoing laser cooling.  A quantitative model for this regime was provided  in our previous work on Rydberg dressed laser cooling \cite{Bounds2018}.

Lastly, we note that in figure \ref{fig:fig6} we observe a factor of $\sim 2$ between the observed atom number and the modelled atom number at the point of plasma termination. We hypothesise that this difference is due to the finite extent of the UCP, which our fits suggest  is less than the size of the atom cloud. Depletion of the atom number due to the plasma therefore occurs preferentially at the centre of the cloud, leaving the wings untouched. This effect is not included in the model. Since we ballistically expand the cloud before imaging, we are unable to study the spatial dependence of the loss. In future experiments it may be possible to directly probe the spatial dependence of the Rydberg excitation in the cloud \cite{Lochead2013}.

In summary, the rate equation model successfully reproduces the plasma formation and decay processes in this complex coupled system. The agreement is good across multiple observables, including the time dependence of the atom number and ion signal and their dependence on cloud density and MOT beam power. The role of ions in the cloud is critical throughout this process, influencing both the seeding of the plasma and the enhancement and partial suppression of Rydberg excitation within the plasma-coupled cloud.

\section{Long-lived Rydberg population}

A final probe of the plasma dynamics is the long-lived Rydberg population observed after the coupling laser is turned off (shown in figure \ref{fig:fig4}). Long-lived Rydberg atoms may form in an UCP due to three-body recombination  or electron-Rydberg collisions. Although previous studies of continuously driven plasmas have found that three-body recombination can modify the plasma dynamics \cite{PhysRevA.89.022701}, we do not expect three-body recombination to be significant at the charge density that we reach \cite{PhysRevLett.86.3759}. Instead we attribute the formation of long-lived Rydberg states to the population of high-angular momentum states by electron-Rydberg collisions, which are significant even at low plasma density \cite{Millen2010}. Once created, these high angular momentum states are slowly ionized by blackbody radiation, giving rise to an ion signal that can be observed several milliseconds after the dressing laser is turned off \cite{Dutta2001}. The long-lived Rydberg population thus contains information about the initial  plasma dynamics even several milliseconds after the plasma disperses. 

\begin{figure}[h]
\centering
\includegraphics[width=14.13cm]{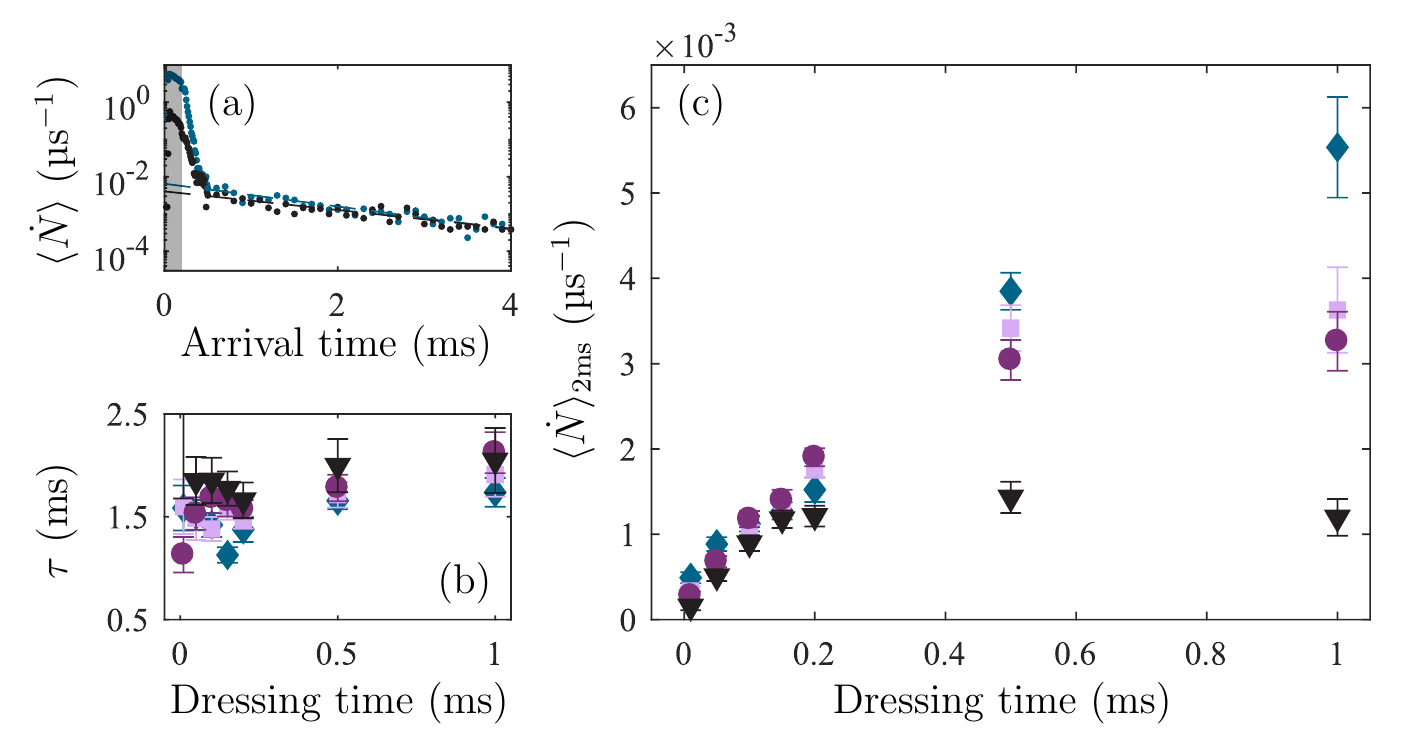}
\caption[Longlived Rydberg atoms]{\label{fig:fig9} (a) Ion detection rate after the coupling laser is turned off for a dressing time of 0.2~ms (indicated by the shading) and atom numbers of 3.4~million (blue) and 0.3~million (black). Fits are shown as dashed lines. Decay lifetimes $\tau$ and ion detection rates 2~ms after dressing begins $\langle \dot{N} \rangle_{2\mathrm{ms}}$ are shown in (b) and (c) respectively for initial atom numbers of 3.4~million (blue diamonds),  1.3~million (pink squares), 1~million (purple circles) and 0.3~million (black triangles). }
\end{figure}

Figure \ref{fig:fig9}(a) shows ion detection rates after the coupling laser has been turned off for the data given in figure \ref{fig:fig6} and taken at  $\mathrm{I_{MOT}} = 9 \ \mathrm{I_{sat}}$ with two different initial atom numbers of 3.4~million and 0.3~million. Despite this order of magnitude difference in atom number, the number of ions detected after the dressing laser is switched off is essentially identical. As illustrated in figure \ref{fig:fig9}(b), the decay rate of the long-lived component of the ion signal is also independent of the atom number and the dressing time. Based on these results, we use the ion detection rate 2~ms after dressing began $\langle \dot{N} \rangle_{2 \mathrm{ms}}$ as a proxy for the total long-lived Rydberg population, and study its variation with dressing time and atom number (figure \ref{fig:fig9}(c)).

The most striking observation in figure \ref{fig:fig9}(c) is that in the first \SI{200}{\micro s} of dressing the rate of creation of long-lived Rydberg atoms is almost independent of atom number.  The ground state atom number varies by an order of magnitude but $\langle \dot{N} \rangle_{2 \mathrm{ms}}$ rises with dressing time at a largely constant rate until the plasma terminates, which occurs sooner for clouds with lower atom number. The electron-Rydberg collision rate is expected to be proportional to the product of the Rydberg population and electron density. Since the plasma is nearly neutral the electron density is related to the ion density less the plasma threshold. The implication is that the fraction of atoms in the Rydberg state must fall with increasing atom number in order to maintain the constant electron-Rydberg collision rate observed figure \ref{fig:fig9}(c). Such a reduction is indeed predicted by our correlated growth model as shown in  figure \ref{fig:fig5}(b); for large ion densities the Rydberg excitation rate falls as the DC Stark shift begins to reduce rather than enhance the Rydberg excitation rate, leading to a cross-over to a Coulomb blockade effect for large enough densities. The long-lived ion signal therefore provides independent confirmation of this cross-over that was  included in the rate equation model that successfully reproduced the formation and decay of the UCP in figures 6--8. In particular it confirms our interpretation of the dependence of the plasma lifetime on the initial atom number in figure \ref{fig:fig8}.

\section{Conclusion}
In summary, we have used  non-destructive and time-resolved ion detection to probe the spontaneous formation of ions in a Rydberg-dressed atomic cloud, and their impact on Rydberg excitation and atom loss. Charges created within the cloud can DC Stark shift nearby atoms into resonance with the excitation laser, leading to strong Rydberg excitation. This Coulomb anti-blockade effect eventually leads to the formation of an UCP, resulting in further ionization and atom loss. The existence of a plasma was evidenced by an ionization threshold and the formation of long-lived high angular momentum Rydberg states. At high density, there is a cross-over to a Coulomb blockade region, where the DC Stark shift from neighbouring ions is large enough to shift nearby atoms out of resonance with the excitation laser. The growth in the ion density with atom number therefore becomes self-limiting.

A key consequence of these results is that careful consideration of the relative sign of the van der Waals interaction and the DC Stark shift is essential in order to minimize charge induced losses in Rydberg dressing experiments. The success of this approach is shown in \cite{Bounds2018}, where the choice of the appropriate Rydberg state led to a significant increase in lifetime. 

More interestingly, the quantitative agreement between our model and the experimental data strongly evidences the ability to tailor the spatial correlation function of Rydberg atoms and ions in the cloud via the Coulomb blockade and anti-blockade mechanisms.  Divalent atoms offer a powerful toolkit of techniques such as autoionization \cite{Millen2010,Lochead2013} and ion imaging \cite{Simien2004,McQuillen2013} that can be used to study these correlations, which may offer new routes to the suppression of disorder-induced heating in plasmas \cite{Bannasch2013}. Coulomb anti-blockade may also be useful to Rydberg-based focused ion beam experiments e.g. \cite{PhysRevLett.115.214802}, enabling higher Rydberg excitation/ion production rates than are allowed due to Rydberg blockade, whilst suppressing disorder-induced heating.

Financial support was provided by EPSRC Grant No. EP/J007021. This project has also received funding from the European Union's Seventh Framework (Grant Agreement No. 612862-HAIRS) and Horizon 2020 (Grant Agreements No. 660028-EXTRYG and No. 640378- RYSQ) research and innovation programmes.

\bibliography{plasma_dressing_bibliography}

\end{document}